\documentclass[english,superscriptaddress,twocolumn,showpacs, longbibliography, aps]{revtex4-1}
\usepackage[T1]{fontenc}
\usepackage[utf8]{inputenc}
\usepackage[active]{srcltx}
\usepackage{verbatim}
\usepackage{amsmath}
\usepackage{graphicx}
\usepackage{subscript}

\makeatletter

\usepackage{amsmath, amssymb} 
\usepackage{braket}
\usepackage{babel}

\providecommand{\abs}[1]{\lvert#1\rvert}

\DeclareMathOperator{\Tr}{Tr}

\usepackage{color}
\usepackage{soul}

\makeatother

\begin{document}
\title{Intra-atomic Hund's exchange interaction determines spin states and energetics of Li-rich layered sulfides for battery applications}

\author{Jae-Hoon Sim}
\affiliation{Centre de Physique Théorique, Ecole Polytechnique, CNRS, Institut
Polytechnique de Paris, 91128 Palaiseau Cedex, France}
\author{D. D. Sarma}
\affiliation{Solid State and Structural Chemistry Unit, Indian Institute of Science,
Bengaluru 560012, India}
\author{Jean-Marie Tarascon}
\affiliation{Coll\`ege de France, 11 place Marcelin Berthelot, 75005 Paris, France}  
\affiliation{Sorbonne Université, Paris, France}
\affiliation{Réseau sur le Stockage Electrochimique de l’Energie (RS2E), Amiens, France}
\author{Silke Biermann}
\affiliation{Centre de Physique Théorique, Ecole Polytechnique, CNRS, Institut
Polytechnique de Paris, 91128 Palaiseau Cedex, France}
\affiliation{Coll\`ege de France, 11 place Marcelin Berthelot, 75005 Paris, France}
\affiliation{Department of Physics, Division of Mathematical Physics, Lund University,
Professorsgatan 1, 22363 Lund, Sweden}
\affiliation{European Theoretical Spectroscopy Facility, 91128 Palaiseau, France,
Europe}

\begin{abstract}
	Motivated by experimental suggestions of anionic redox processes helping to design higher energy lithium ion-battery cathode materials, we investigate this effect using first-principles electronic structure calculations for Li-rich layered sulfides.
	We identify the determination of the energetic contribution of intra-atomic Hund's exchange coupling as a major obstacle to a reliable theoretical description. We overcome this challenge by developing a particularly efficient flavor of charge-self-consistent combined density functional + dynamical mean-field theory (DFT+DMFT) calculations.
 Our scheme allows us to describe the spin ground states of the transition metal $d$ shell, the electronic structure of the materials, and its energetics.
 As a result of the high-spin to low-spin transition the average intercalation voltage shows intriguing non-monotonic behavior.
	We rationalize these findings by an analysis of the fluctuations of spin and charge degrees of freedom. 
	Our work demonstrates the relevance of most recent insights into correlated electron materials for the physics of functional materials such as Li-ion battery compounds.
\end{abstract}
\date{\today}
\maketitle

\section{Introduction}

Many modern technologies such as mobile devices or electric cars hinge upon the development of high-energy density batteries. 
Finding cathode materials fulfilling the mandatory criteria of being safe and inexpensive with high capacity and intercalation voltage is a bottleneck in the field.
A traditional, extensively studied and commercially successful example is LiCoO\textsubscript{2} \cite{MizushimaGoodenough_Li1980}.
While for many years, the main strategy for cathode materials in Li-ion batteries purely relied on cationic redox, lithium-rich layered manganese oxides have recently attracted great interest, involving both cationic and anionic redox processes \cite{LuDahn_Layered2001, AssatTarascon_Fundamental2018,LiXia_Anionic2017}.
The lithium-rich layered manganese oxide Li$_{2}$MnO$_{3}$ is the parent compound of the currently used Li{[}Li$_{x}$Mn$_{y}$Ni$_{z}$Co$_{1-x-y-z}${]}O$_{2}$, which are famous for their reversible high capacities exceeding 250 mAh/g \cite{LiXia_Anionic2017, AssatTarascon_Fundamental2018, ThackerayHackney_Li2MnO3stabilized2007}. Undesired consequences of anionic redox processes are, however, potential capacity loss and structural degradation, as well as hysteresis.

Recently, new Li-rich layered sulfides Li$_{x}${[}Li$_{0.33-2y/3}$Ti$_{0.67-y/3}$Fe$_{y}${]}S$_{2}$
have been reported \cite{SahaTarascon_Exploring2019}. Its negligible cycle irreversibility, mitigated
voltage fade upon long cycling, low voltage hysteresis, and fast kinetics
suggest a new direction to alleviate the practical limitations of using the anionic redox mechanism.
Motivated by their experimental realization, we perform
first-principles calculations for the above-mentioned Li-rich layered sulfides, using as prototypes the fully lithiated and delithiated materials. We focus on the $y=1/3$ case, i.e., Li$_{x+0.11}$Ti$_{0.56}$Fe$_{0.33}$S$_{2}$
(LTFS$_{x}$). Note that the value $y=1/3$ is close to the optimal value
reported in experiment, namely $y=0.3$ \cite{SahaTarascon_Exploring2019}.

First-principles electronic structure calculations based on density
functional theory (DFT) within the local density approximation (LDA)
and generalized gradient approximation (GGA) have developed into a tremendously useful tool for addressing materials properties and even helping design functional materials.
Nevertheless, these approximations have known limitations in  describing materials containing open $d$- or $f$-shells with sizable Coulomb interactions.
Indeed, the standard LDA/GGA description based on the delocalized electron gas picture is not an optimal starting point for the rather localized behavior of $d$- or $f$-electrons. In the materials of interest here, as we will see below, a particular challenge is the necessity of capturing the spin-state of these localized electrons correctly, since the energetics of the materials depend on the corresponding exchange contribution.
A second limitation arises for properties that involve excited states, which are in principle inaccessible to standard DFT, even if the exact ground state energy functional was available.
Over the last decades, combinations of LDA/GGA with many-body techniques have evolved into extremely useful and well-established tools overcoming these limitations.
Among these, DFT plus dynamical mean-field theory (DFT+DMFT) \cite{MetznerVollhardt_Correlated1989, GeorgesKotliar_Hubbard1992, ZhangKotliar_Mott1993, AnisimovKotliar_Firstprinciples1997, LichtensteinKatsnelson_Initio1998} and DFT$+U$ \cite{AnisimovAndersen_Band1991, LiechtensteinZaanen_Densityfunctional1995} stand out as methods of choice.

Here, we suggest a new flavor of the charge-self-consistency in DFT+DMFT methods, to describe the observed electronic ground states and the competing phases of our target materials, namely a high-spin (HS) Mott insulator, a correlated metal, and a low-spin (LS) band insulator phase.
We show that the energetics of the cathode materials largely depends on the spin ground state, revealing that the impact of the Hund's coupling is not only of theoretical interest but has practical implications for battery materials. 
These findings place our target materials in the context of the nowadays celebrated ``Hund's materials'', where intra-atomic exchange amplifies the consequences of strong local Coulomb interactions \cite{deMediciGeorges_JanusFaced2011, GeorgesMravlje_Strong2013}.
They should be viewed as strongly correlated electron systems, the properties of which can be described in the language of modern first-principles many-body techniques.

The paper is organized as follows. In Sec. \ref{sec:Methods-and-formalism}, we present details of our calculational techniques. In particular, we have developed a new flavor for DFT+DMFT that takes into account charge redistributions due to correlation effects. In Sec. \ref{sec:Results-and-discussion}, the results of our calculations are presented and comparisons between static (DFT$+U$) and dynamic (DFT+DMFT) mean field approximations to electronic Coulomb correlations are made. Finally, a summary and some conclusions are given in Sec. \ref{sec:Conclusion}.
\section{Methods and formalism: charge-self-consistent DMFT scheme \label{sec:Methods-and-formalism}}

Within the DFT+DMFT framework, the charge-density $\rho (r)$ and the local Green's function $G_{\rm loc}(i\omega_n)$ are determined by the stationary condition of the free energy functional $\Gamma[\rho, G_{\rm loc}]$ \cite{GeorgesRozenberg_Dynamical1996}.  Single-shot (i.e. non-charge-self-consistent) DFT+DMFT calculations are performed in many works in the literature, where $\rho$ is given from LDA/GGA calculations.
In the case of the LTFS, however,
due to the larger covalency between S-$p$ and Fe-$d$ compared
to oxides, it is important to take into account the effect of the charge
redistribution due to correlation effects. The importance of the
charge redistribution is demonstrated by the fact that conventional
one-shot DFT+DMFT with fixed DFT charge density predicts a significantly underestimated voltage as we will discuss in Sec. \ref{sec:Results-and-discussion}.
On the other hand, given the large unit cells with low symmetry,
fully charge-self-consistent DFT+DMFT calculations are hardly
accessible due to the computational cost.

In this paper, we use an
efficient way to calculate the total energy given as
\begin{align}
	E_{{\rm DFT}+{\rm DMFT}}= & E_{{\rm GGA}}[\rho_{U}]-\braket{H_{{\rm KS}}^{GGA}[\rho_{U}]}_{\rho_{U}}+\Tr(H_{{\rm KS}}[\rho_{U}]G)\nonumber \\
	& +\frac{1}{2}\Tr\Sigma G-E_{{\rm DC}}^{{\rm FLLnS}},\label{eq:E_DMFT}
\end{align}
where the GGA energy functional $E_{{\rm GGA}}[\rho_U]$ and  $H_{KS}^{GGA}[\rho_U]=\frac{\delta E_{{\rm GGA}}}{\delta\rho(r)}\bigr|_{\rho = \rho_U}$
are evaluated at the ground state density $\rho_{U}(r)$  obtained from GGA$+U$ calculations.
The double counting $E_{{\rm DC}}^{{\rm FLLnS}}$
is taken to be the spin-independent "fully localized limit" double counting using the spin-averaged occupations \cite{AnisimovSawatzky_Densityfunctional1993}, dubbed "FLL-nS" in Ref. \cite{YlvisakerKoepernik_Anisotropy2009}.

We stress that $\rho_{U}$ differs from the ground state
density $\rho_{0}$ of $H_{{\rm KS}}[\rho_{U}]$ due to the absence
of the $+U$ contribution in $E_{{\rm GGA}}$ and $H_{KS}^{GGA}$.
One can understand our scheme as follows: i) First, charge-self-consistent
DFT+DMFT calculations are performed with a Hartree-Fock impurity
solver. ii) Given the resulting charge density, we produce the Hamiltonian
$H_{{\rm HK}}+H_{{\rm int}}$ on which the DMFT calculations are performed
with a more sophisticated impurity solver, namely continuous-time
quantum Monte Carlo (CT-QMC) simulations \cite{WernerMillis_ContinuousTime2006,WernerMillis_Hybridization2006}.
This approach is not only computationally efficient, but also useful
for comparing the results to the DFT$+U$ calculations. Because both
DFT$+U$ and +DMFT use the same charge density $\rho_{U}$, it is
easier to investigate the correction stemming from the dynamical self-energy, without ambiguity. Once we use the HF solver in step ii), for example, $E_{{\rm DFT+DMFT}}$ exactly reduces to the DFT$+U$ results.


In Eq. (\ref{eq:E_DMFT}), we have a adopted charge-only-dependent exchange-correlation GGA functional and FLL-nS double counting. The FLL-nS double counting scheme has been introduced within the DFT$+U$ functional \cite{YlvisakerKoepernik_Anisotropy2009} (cFLL in the notation of Ref. \onlinecite{RyeeHan_Effect2018}):
\begin{equation}
	E_{{\rm DFT}+U}[\rho_{U}]=E_{{\rm DFT}}[\rho_{U}]+E_{{\rm int}}^{U}[n_{\alpha\beta}^{\sigma}]-E_{{\rm DC}}^{{\rm FLLnS}}[N_{d}],\label{eq:TotEU}
\end{equation}
where $E_{{\rm int}}^{U}$, and $E_{{\rm DC}}^{{\rm FLLnS}}$ are the Hartree-Fock energy of the local Coulomb
interaction, and the double counting contribution, respectively.
We note that the $E_{{\rm GGA}}$
and $E_{{\rm DC}}^{{\rm FLLnS}}$ contributions to the total energy
depend only on the charge density $\rho=\rho_{\uparrow}+\rho_{\downarrow}$
and the occupation of the $d$-shell $N_{d}=\sum_{\sigma,\alpha}n_{\alpha\alpha}^{\sigma}$
, respectively. The spin density contributions to the total energy
are controlled solely by the interaction term $E^{U}_{{\rm int}}$ or by the self-energy $\Sigma^\sigma$ in the DMFT formalism. For
more details, see Appendix \ref{sec:On-site-Coulomb-energy}.

Benchmarks of different DFT$+U$ formalisms \cite{AnisimovAndersen_Band1991, AnisimovSawatzky_Densityfunctional1993, DudarevSutton_Electronenergyloss1998, LiechtensteinZaanen_Densityfunctional1995}
have been performed in previous studies \cite{YlvisakerKoepernik_Anisotropy2009,RyeeHan_Effect2018}.
The spin-dependent DFT$+U$  extensions exhibit
seemingly unphysical behavior, including an unreasonable $J$ dependence
of structural parameters in nickelates \cite{PavariniInstituteforAdvancedSimulation_LDA2011}
and the energy difference between high-spin/low-spin states in 3$d$ and 4$d$ transition-metal compounds \cite{RyeeHan_Effect2018}.
Similar arguments were also given in the DFT+DMFT context \cite{ChenMarianetti_Density2015,ParkMarianetti_Total2014,ChenMillis_Spindensity2016}.

\section{Results and discussion\label{sec:Results-and-discussion}}

\subsection{Electronic structure of the layered sulfides within GGA and GGA$+U$ }

\begin{figure}
	\includegraphics[width=1\columnwidth]{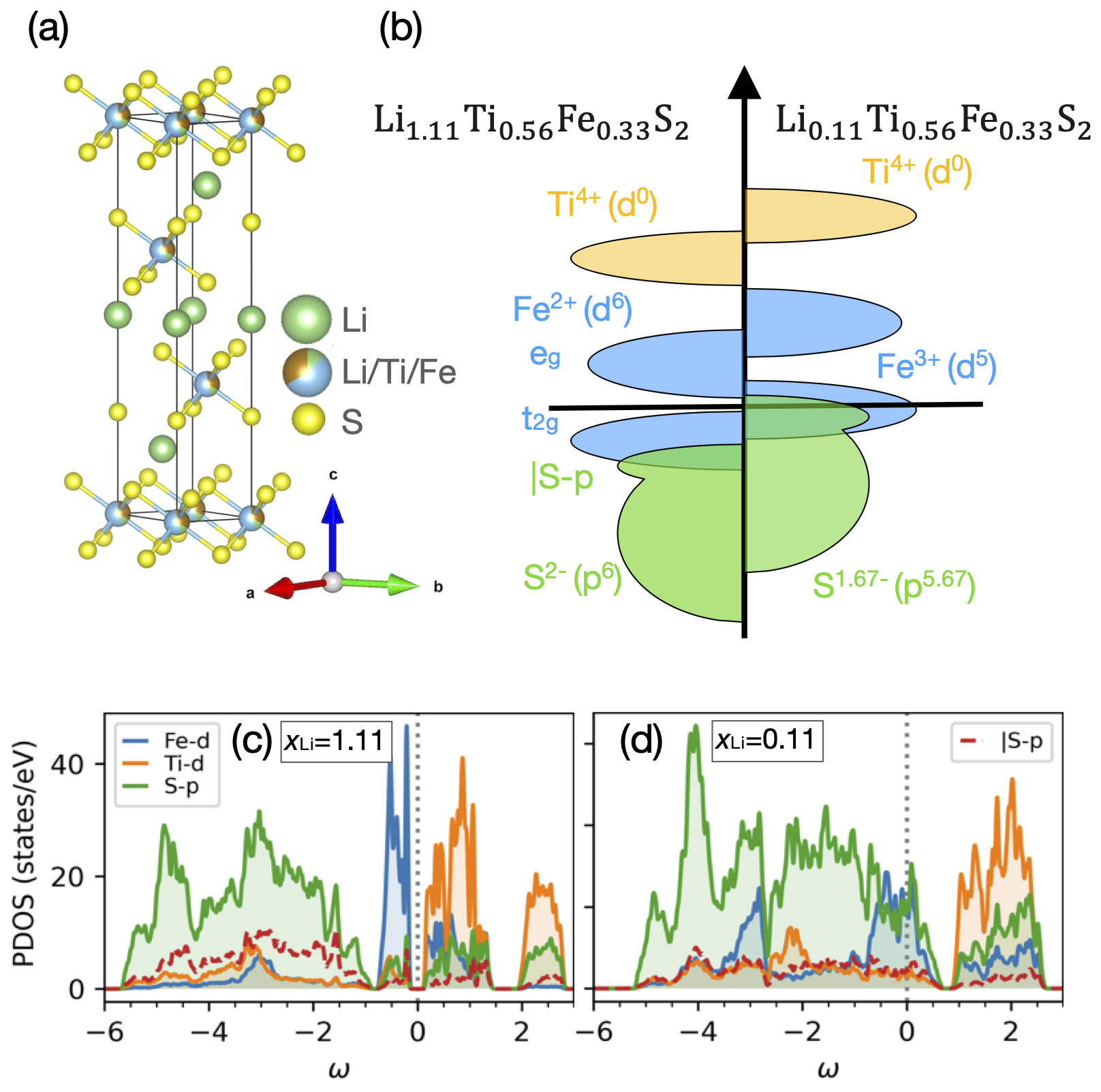}
	\caption{
		(a) Crystal structure of Li$_{x}${[}Li$_{0.33-2y/3}$Ti$_{0.67-y/3}$Fe$_{y}${]}S$_{2}$. Green, blue, and yellow spheres represent Li ions, a mixture of Li/Ti/Fe ions, and S ions, respectively. Visualization was conducted using VESTA \cite{MommaIzumi_VESTA32011} (b) Schematic electronic structure near the Fermi level for the $y=1/3$ case. |S-p denotes non-bonding S $3p$ states. The oxidation states are estimated from simple charge counting.
		(c, d) The projected density of states (PDOS) obtained from  GGA calculations. Red dashed lines represent the S $3p$ contained in the octahedron with Li as a central atom, capturing the non-bonding states. 
		 \label{fig:Schematic}
	}
\end{figure}

	The Li-rich layered Li$_{1.33-2y/3}$Ti$_{0.67-y/3}$Fe$_{y}$S$_{2}$ can be thought of as a Li$_{1.33}$Ti$_{0.67}$S$_{2}$ compound, where some of the Ti$^{4+}$ have been substituted 
	by Fe$^{2+}$ with a charge compensation by adjusting the Li content.
	The crystal structure is shown in Fig. \ref{fig:Schematic}(a) \cite{SahaTarascon_Exploring2019}. It consists of stacks of transition metal disulfide layers crystallizing in a honeycomb lattice structure. Upon lithiation, Li-ions intercalate in the interlayer space or within the layers.
At the DFT level, the Fermi surface is dominated by Fe$^{2+}$ $d$-derived bands, which are strongly hybridized with S$^{2-}$ $p$ orbitals. The Ti$^{4+}$ ions are electronically inactive because of their $3d^0$ configuration. Upon removal of Li, both cationic Fe$^{2+}$ and anionic S$^{2-}$ contribute to the corresponding redox processes \cite{SahaTarascon_Exploring2019}. 
As shown  in Ref. \onlinecite{SahaTarascon_Exploring2019}, the experimentally observed capacity of 245 mAhg$^{-1}$ for   Li$_{1.13}$Ti$_{0.57}$ Fe$_{0.3}$S$_2$  ($y = 0.3$), which corresponds to the removal of $\sim 1.06$ Li ions per formula unit, cannot be reached by cationic redox alone, even when assuming a multi-electron oxidation from Fe$^{2+}$ to Fe$^{4+}$. 
Hence anionic redox processes, leaving unoccupied sulfur $p$ states near the Fermi level in Fig. \ref{fig:Schematic}(d), have to be invoked to understand the high capacity of these materials.
From a simple charge count, we can estimate the oxidation states Fe$^{3+}$ ($d^5$) and S$^{1.67-}$  ($p^{5.67}$) 
for fully charged (i.e., fully oxidized) LTFS$_0$ (see Fig. \ref{fig:Schematic}(b)).

Fig. \ref{fig:Schematic}(c) and (d) show the projected density of states (PDOS) corresponding to the non-spin-polarized Kohn-Sham band structure obtained from GGA calculations. 
Within the GGA, fully discharged LTFS$_{1}$ is a band insulator with fully occupied $t_{2g}$ orbitals,
while LTFS$_{0}$ is metallic with nominal $t_{2g}^{5}e_{g}^{0}$ configuration.
Although the FeS\textsubscript{6} octahedra are slightly distorted due to the different size of the Li, Ti, and Fe ions in the
metal layer, we will refer to the three lower-lying orbitals as $t_{2g}$ and to the two higher-lying ones as $e_{g}$. 
Due to the large hybridization between the Fe-$d$ and S-$p$ states in LTFS$_{0}$, larger ligand field splittings are expected from different hopping strengths for $t_{2g}$ and $e_g$ orbitals.
To calculate the ligand field splittings, projective Wannier functions
within an energy window of $W=[-1,3]$ eV were constructed.
The on-site energy level differences between the three lower- and two higher-lying orbitals, denoting the crystal-field splitting, amount to 1.04 eV and
1.78 eV for LTFS$_{1}$ and LTFS$_{0}$, respectively.
Here the smaller crystal-field splitting for LTFS$_1$ compared to LTFS$_0$ also implies a redox-driven spin-state transition and is consistent
with the experimentally observed high-spin and low-spin states for  LTFS$_1$ and LTFS$_0$, respectively \cite{WatanabeYamada_RedoxDriven2019, SahaTarascon_Exploring2019}.

 The red dashed lines in Fig. \ref{fig:Schematic} (c, d)  represent the S $3p$ states of the S atoms forming the LiS$_6$ octahedra, which are of non-bonding nature. 
We calculated $r_{nb}$, the fraction of the partial charge residing in the ``non-bonding'' state by integrating the relevant partial DOS (red dashed and solid green lines) shown in Fig. 1 (c) and (d) within a given energy window.
For LTFS$_1$, $r_{nb}$ is 0.80 in the energy window $[-2, 0]$, while 0.46 for $[-6,-2]$. The larger $r_{nb}$ near the Fermi level clearly shows the non-bonding contribution in the above bonding states as presented in Fig. \ref{fig:Schematic}(b).

\begin{figure}
\includegraphics[width=1\columnwidth]{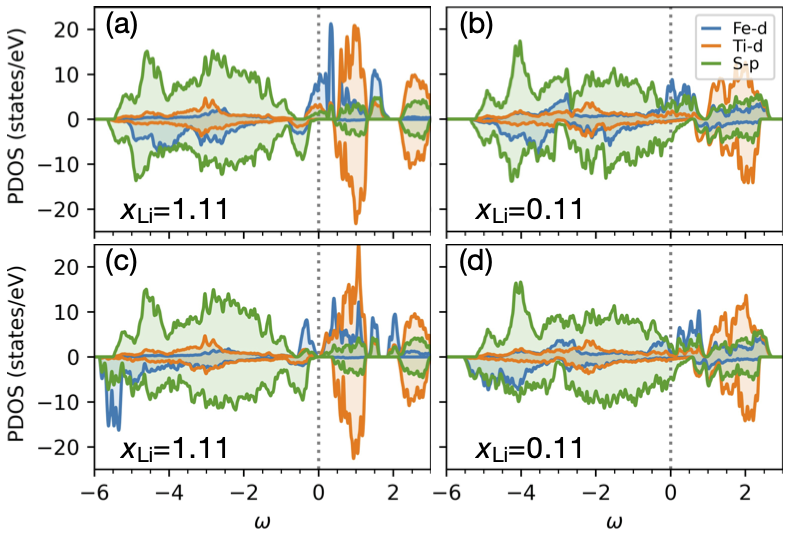}\caption{The projected density of states (PDOS) calculated from (a,b) spin-polarized GGA (SGGA), and (c, d) GGA$+U$. The upper and
lower panels in the PDOS plots represent up and down spin components,
respectively. \label{fig:PDOS-GGA}}
\end{figure}

Fig. \ref{fig:PDOS-GGA}(a) and (b) present the PDOS with spin-polarized
GGA (SGGA). LTFS$_{1}$ is metallic with HS Fe-$d$ configuration,
while the Fe-$d$ shell in the 
 LTFS$_{0}$ is not fully polarized with partially unoccupied
minority spin components.
The spin-configuration is consistent with the previously reported assignments based on Mössbauer spectroscopy experiments, where the fitted spectra show finite quadrupole splitting for both Fe$^{3+}$ and Fe$^{2+}$ oxidation states \cite{SahaTarascon_Exploring2019}.
Interestingly, when creating a hypothetical oxide compound by 
replacing S by O, one can artificially tune the ligand splitting due to the corresponding smaller overlap between O-$p$ and Fe-$d$ orbitals as compared to the one of S-$p$ and Fe-$d$ (not shown). The result is a HS configuration
for both end-member compounds with large spin magnetic moments on the Fe sites,
$M_{s}=3.80$ and $3.49$ $\mu_{B}$, respectively.
The appearance of these different spin configurations implies that a change of the anion from O to S can significantly affect battery properties such as the capacity  and more so  the operating voltage\cite{SahaTarascon_Exploring2019, WatanabeYamada_RedoxDriven2019}.

The GGA$+U$ results for LTFS$_1$ and LTFS$_0$ are shown in Fig. \ref{fig:PDOS-GGA}(c, d), respectively.
The effect of the Coulomb interaction between the localized $d$-electrons
is clearly seen from the Mott insulating gap for LTFS$_{1}$ and from the suppression of the Fe-$d$ contribution near the Fermi-level.
Here, the insulating gap opening is driven by a reduction in the occupancy fluctuations over the three non-degenerate low-lying orbitals, referred to as $t_{2g}$ for convenience, due to the energy penalty of the $\frac{1}{2}(U-J)\sum_{\alpha\sigma}(n_{\alpha\sigma}-n_{\alpha\sigma}^{2})$ term in $\Delta E^U$ (See Appendix \ref{sec:On-site-Coulomb-energy}). From the SGGA calculation, for example, the eigenvalues of the density matrix for the $t_{2g}$ subspace of the minority spin component are given as 0.27, 0.30, and 0.57, while GGA$+U$ provides 0.26, 0.25, and 0.83.
For LTFS$_0$, 
however, the spin magnetic moments, $M_S = 2.30\:\mu_B$ are enhanced from the result of GGA, $M_S=1.84\:\mu_B$, due to the $-\frac{J}{4}M_S^{2}$ contribution in  $\Delta E^U$.
Considering the experimental
assignment of the LS state to Fe$^{3+}(d^{5})$ ions in the fully charged
sample and the ideal magnetic moments $\sim1$ $\mu_{B}$ of the $d^{5}$
in the LS state \cite{SahaTarascon_Exploring2019},
the predicted magnetic moments by the DFT+$U$ calculations are significantly overestimated, stressing the limitations of the static mean field approximation done in GGA+$U$.
This motivated us to go beyond GGA+$U$ and perform GGA+DMFT calculations for these same materials.
Detailed comparisons are presented in the next section.

\subsection{Electronic structure of the layered sulfides within GGA+DMFT}

\begin{figure}
\includegraphics[width=1\columnwidth]{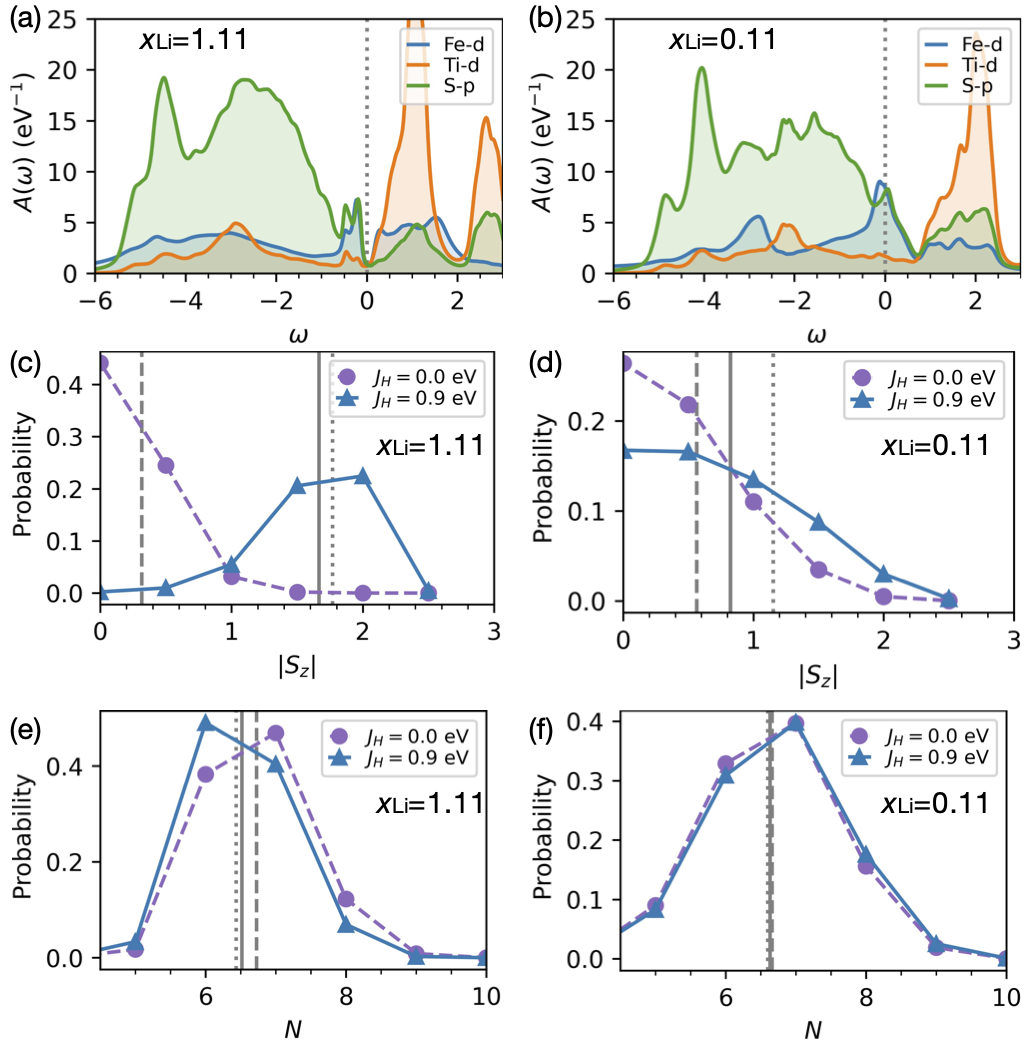}\caption{(a, b) The spectral function from GGA+DMFT. Probability of the atomic
states in Fe-$d$ shell projected onto the (c, d) spin $S_{z}$ and
(e, f) number of electrons $N$. 
The solid and dashed vertical lines indicate the averaged value for $J_H = 0.9$ and $0.0$ eV, within GGA+DMFT and the dotted line from the GGA$+U$ calculations.
\label{fig:PDOS-U}}
\end{figure}

To capture the electronic ground state or quasi-particle electronic structure of the HS Mott insulating LTFS$_1$ and the LS correlated metallic LTFS$_0$, in this section, we have employed GGA+DMFT.
Specifically, we study the effect of dynamic correlations between the Fe-$d$ electrons on the electronic structure and the magnetic properties
using the Hubbard interaction $U$ and the Hund's exchange coupling $J_H$ as parameters.
The paramagnetic GGA+DMFT calculations are performed at an inverse temperature of $\beta=40$ eV$^{-1}$ corresponding to about 290 K.
Since the occupied valence orbitals
are drastically modulated by spin-state transition,
the correlation between spin states and redox properties is expected.

In Fig. \ref{fig:PDOS-U}(a) and (b),
we present orbital-resolved spectral functions from paramagnetic GGA+DMFT calculations. The difference of these results to the DOS from the static GGA$+U$ approximation is significant, in particular for LTFS$_{0}$ near the Fermi-level, underlining the importance of the dynamic correlation effects. 
Within GGA$+U$, the Fe-$d$ contribution to the DOS near the
Fermi level is small, contrasting to the large Fe-$d$ character in the GGA+DMFT result.
The suppression of the Fe-$d$ contribution in the GGA$+U$ calculation
originates from the suppression of charge fluctuations by the Coulomb interactions (see energy penalty term $\Delta E^U$ in Appendix \ref{sec:On-site-Coulomb-energy}), favoring integer occupation.

The reason why GGA$+U$ struggles to capture the electronic structure
of the correlated metallic phase can be traced back to the absence of dynamical fluctuations in this method. This can be seen from the probability distribution of the atomic configurations. Indeed, solving the DMFT by Monte Carlo sampling has the advantage to give direct access to the charge and spin fluctuations: Fig. \ref{fig:PDOS-U}(c-f)
shows the probabilities of the different atomic configurations in terms of the magnitude of the spin component and the
orbital occupancies. As expected, a sharp peak at around $\abs{S_{z}}=2$ is observed in Fig. \ref{fig:PDOS-U}(c)
indicating the HS configuration. For metallic LTFS$_{0}$, the spin
fluctuations $\delta\braket{|S_{z}|}^{2}=0.33$ with mean value of
spin magnetic moments $\braket{\abs{S_{z}}}_{{\rm DMFT}}=0.82$ are lager than that of LTFS$_{1}$ ($\delta\braket{|S_{z}|}^{2}=0.15$ and $\braket{\abs{S_{z}}}_{{\rm DMFT}}=1.67$).
We note that these fluctuations of the spin magnetic moments on the
Fe sites cannot be captured in a static-mean-field theory such as DFT+$U$.
The calculated magnetic moments in metallic LTFS$_{0}$ compounds, $M_S=2\braket{\abs{S_{z}}}_{\rm DMFT}\sim1.65$ $\mu_{B}$, are in better agreement with the experimental assignment of the LS state  \cite{SahaTarascon_Exploring2019} than GGA+$U$ (2.30 $\mu_B$).

The probability distribution projected onto the number of electrons is also presented in Fig. \ref{fig:PDOS-U}(e,
f). Both compounds show non-negligible charge fluctuations with $\delta\braket{N}^{2}=0.48$
and $0.93$ for LTFS$_{1}$ and LTFS$_{0}$, respectively. As expected, the metallic LTFS$_{0}$ shows larger fluctuations
than insulating LTFS$_{1}$.
Due to the strong hybridization with the S-$p$ orbitals
the mean occupation is slightly larger than nominally expected.
In particular, the electron occupancy of 6.63 of the Fe-$d$ orbitals
for LTFS$_{0}$ is larger than that of LTFS$_{1}$, 6.52,
which is consistent with stronger $d$-$p$ hybridization as noticed
in Fig.\ref{fig:PDOS-GGA}(b). On the other hand, 0.67 electrons are depleted from the sulfur ion coordinated with four Li and two Fe/Ti ions, while 0.52 electrons come from other sulfur ions surrounded by six transition-metal ions.
The strong intermixing of the Fe-$d$ and
S-$2p$ states visible from Fig. \ref{fig:PDOS-U}(b), corresponding to holes created in the S-p states, reveals the covalent nature of Fe-S bonding.
The different numbers of holes in the states associated with the different sulfur
ions are a proxy for the different chemical environments, revealing distinct mechanisms occuring when extracting
electrons from unhybridized S $2p$ non-bonding states sitting in
Li--S--Li configurations \cite{SeoCeder_Structural2016,SahaTarascon_Exploring2019}.

\begin{figure}
\includegraphics[width=1.0\columnwidth]{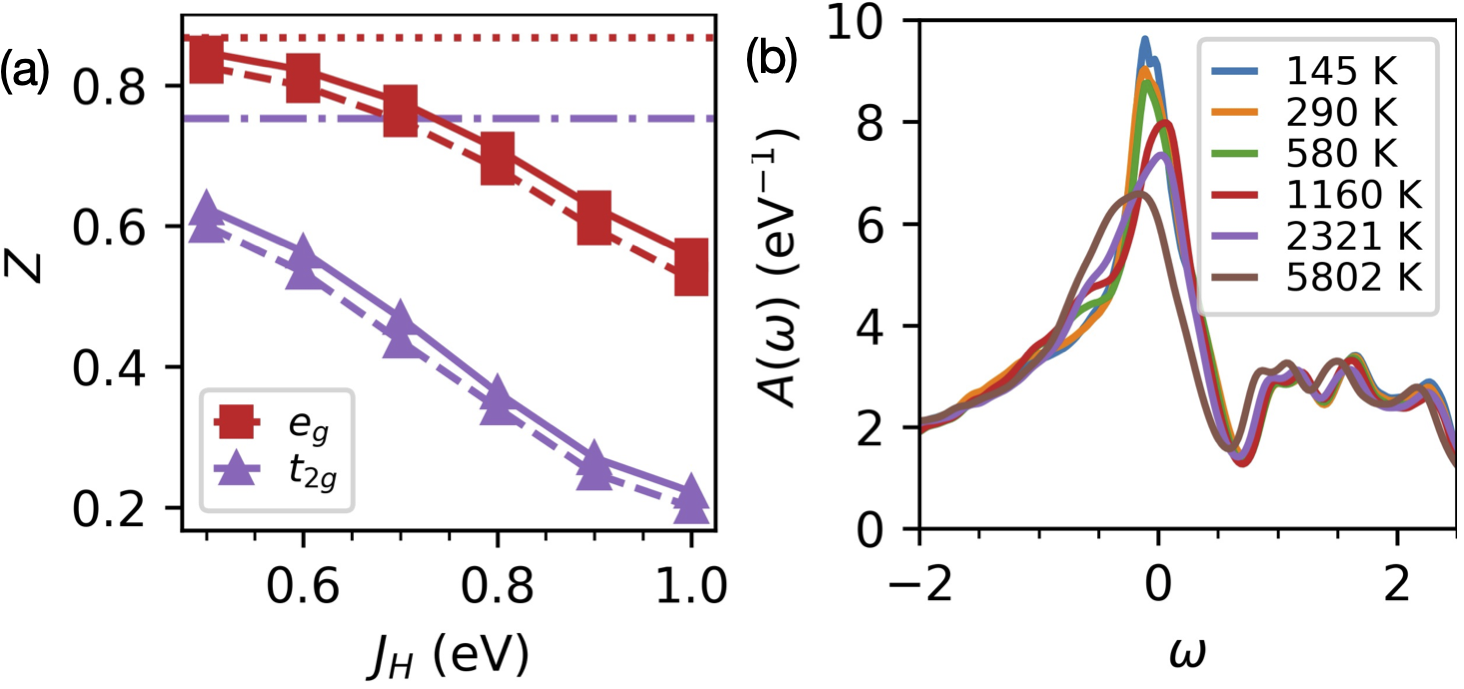}

\caption{(a) Quasiparticle weight $Z$ of the $t_{2g}$ and $e_{g}$ orbitals
are shown as functions of Hund's coupling $J_{H}$ for LTFS$_{0}$
with $U=3.0$ (solid line) and $3.5$ eV (dashed line). The system
gradually evolves to a more strongly correlated state with increasing $J_{H}$,
while it seems weakly correlated at $J_{H}=0$ with $Z_{t_{2g}}(Z_{e_{g}})=0.75(0.8$7)
(dash-dot and dotted line, respectively) 
(b) Spectral functions projected onto the Fe-$d$ orbitals calculated for different temperatures.
 \label{fig:Quasi-particle-weight}}
\end{figure}

In metallic systems, dynamic correlation effects can be quantified
by means of the quasi-particle renormalization factor $Z\approx\left[{\rm 1-Im}\Sigma(i\omega_{0})/\omega_{0}\right]^{-1}$,
where $\Sigma(i\omega_0)$ is the electronic many-body self-energy calculated from DMFT at the first Matsubara frequency. This factor is close to unity for a weakly correlated system, while a small $Z$ indicates a strongly correlated phase. In our case, LTFS$_{1}$ is a Mott insulator, albeit with a quite characteristic many-body behavior.
For LTFS$_{0}$, the calculated value of $Z$ is 0.26 for the $t_{2g}$
orbitals, demonstrating the presence of rather strong correlations even in the metallic phase of this system.
This rather strong renormalization of the quasi-particle states in this system stems from the intra-atomic Hund's exchange coupling $J_{H}$, making LTFS$_{0}$ a realization of what in the literature is sometimes called a ``Hund's metal''
\cite{deMediciGeorges_JanusFaced2011, GeorgesMravlje_Strong2013}, as demonstrated in Fig. \ref{fig:Quasi-particle-weight}(a).
In this figure, the quasiparticle
weight $Z$ as a function of $J_{H}$ is shown for two different interaction strengths, $U=3.0$ and $3.5$
eV. We observe that $Z$ diminishes gradually as $J_{H}$ or $U$
increase with a more pronounced dependency on $J_H$.
The fact that neither GGA nor GGA$+U$ can capture the electronic ground state or quasi-particle electronic structure
and the strong quasiparticle renormalization evidenced by the small
$Z$ factor in the metallic LTFS$_{0}$ phase allows us to attribute the system to the class of strongly correlated materials.
To investigate the nature of the correlations
in LTFS$_{0}$, we study in more detail the spectra projected onto the Fe-$d$ orbitals,
up to high temperature (for fixed lattice constant and atomic positions)  \cite{DengKotliar_Signatures2019}. 
The spectra shown in Fig. \ref{fig:Quasi-particle-weight}(b) are characterized by two distinct peaks, namely $t_{2g}$ states near the Fermi level and  $e_g$ states around an energy of 1.5 eV.
As temperature increases, the $t_{2g}$ bands gradually broaden, while their overall shape remains unchanged. This is a distinctive feature from the traditional correlated metal which is near the Mott transition, where two broad side peaks, the so-called Hubbard bands, with gap features develop at high temperatures \cite{DengKotliar_Signatures2019}.

\subsection{\textmd{\normalsize{}Average intercalation voltage\label{subsec:Average-intercalation-voltage}}}

\begin{figure}
\includegraphics[width=0.85\columnwidth]{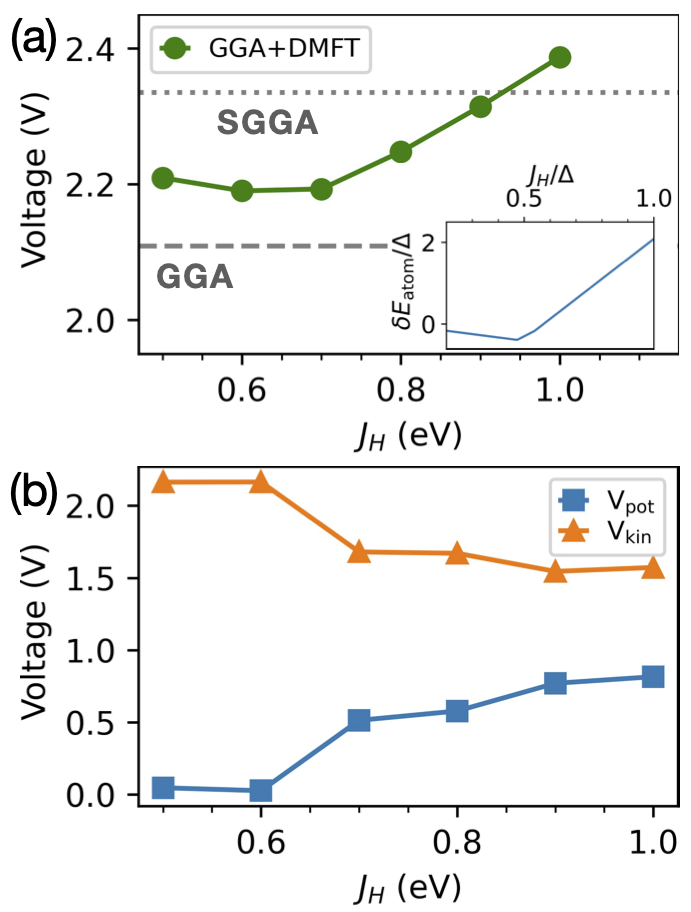}

\caption{(a) Calculated intercalation voltage as a function of $J_{H}$ obtained
from GGA+DMFT (green). The dashed and dotted horizontal
lines indicate the calculated voltage via GGA and spin-polarized GGA. (inset) Estimated voltage in atomic limit. See main text for more details.  (b) Voltage decomposed into Coulomb interaction energy and kinetic energy contribution.   \label{fig:Calculated-intercalation-voltage}}
\end{figure}

We now turn to another key quantity of a redox couple of battery materials, the intercalation voltage, which for our systems has been determined experimentally to be around 2.5 eV \cite{SahaTarascon_Exploring2019}.
In theoretical calculations, the average battery voltage can be evaluated as $V=\left(E({\rm LTFS_{0}})+E({\rm Li})-E({\rm LTFS_{1}})\right)/e$, where the three terms are the total energies of the delithiated system, elemental Li and the lithiated system. 
We have calculated this quantity using different total energy functionals. The conventional
GGA results, 2.11 V, underestimate the experimental value by around $0.37$
eV \cite{SahaTarascon_Exploring2019}, while the SGGA result $V=2.34$ V is in better
agreement with experiment. 
An accurate description
of the spin state of the Fe-$d$ shell is important not only for the
electronic excitations, but also for the energetics, as seen from the
difference between GGA and SGGA. The relatively small error
of SGGA (without $U$) for this layered sulfide indicates that the strong local
Coulomb interactions among the Fe-$d$ electrons are largely screened
by the S-$p$ orbitals, having covalent character.

Let us now discuss the impact of the local correlations that govern the behavior of the Fe-$d$ orbitals.
Within GGA+DMFT, the calculated voltage is 2.31 V, which is close to the conventional SGGA result.
It exhibits only a weak dependence on the Hubbard $U$; for example, $V=$ 2.34 V 
within GGA+DMFT calculations with $U=3.5$, confirming
that the on-site Coulomb interactions are largely screened due to
the large Fe-$d$ and S-$p$ orbital hybridisations. However, we will see that the
spin state of the valence orbitals, which is determined by $J_{H}$ and the
ligand field splitting, are important to describe the properties of the
cathode material.

We now turn to the effect of Hund's coupling $J_{H}$ and the spin-states
on the energetics e.g., the operating voltage of the battery. As can be seen from Fig. \ref{fig:Calculated-intercalation-voltage}(a), the voltage
curve as a function of Hund's $J_{H}$, while keeping $U=3.0$ eV fixed, shows a V-shape behavior. For small values of the Hund's exchange $J_{H}$, where LTFS$_{1}$ is in the
LS state, the voltage decreases at a rate of 0.01 V per 0.1 eV change in
$J_{H}$, while in the regime of large $J_{H}$ where  LTFS\textsubscript{1} is in the
HS phase, the voltage starts to increase by 0.07 V per 0.1 eV $J_{H}$.
This behavior can be explained in a simple manner by considering the atomic limit.
Assuming an isolated Fe-$d$ shell with nominal electron configuration, one easily obtains (see Appendix B) the  V-shape voltage curve as a function of $J_H$:
\begin{align}
	\delta E_{\rm atom} &=E(d^{6})-E(d^{5})   \nonumber  \\
	                                &\propto 
	                                 \begin{cases}
	                                 	4.89J_{H} - 2\Delta,  & \text{for }J_{H}/\Delta>0.35\\
	                                 	-0.82J_{H}, & \text{for }J_{H}/\Delta<0.35.
                                 	\end{cases}
\end{align}
As shown in the inset of Fig. \ref{fig:Calculated-intercalation-voltage}(a), the overall trends, including the V-shape behavior and the larger (smaller) slope in the HS (LS) region, are well described in this simple atomic model.
Further details can be found in Appendix \ref{sec:On-site-Coulomb-energy}.
 Qualitative differences are observed in the voltage for the LS and HS cases. Fig. \ref{fig:Calculated-intercalation-voltage}(b) shows the operating voltage decomposed into 
 	the Coulomb interaction $V_{\rm pot}$, where the Hartree and exchange-correlation energies corrected by $+U$ terms are taken into account, and the kinetic energy $V_{\rm kin}$ contribution.  
 For $J_H>0.7$ it is clearly seen that the increase in voltage is attributed to the $V_{\rm pot}$ part. In other words, a proper theoretical description of the energy gain from Hund's interactions in the HS state is a decisive element to capture the operating voltage.

The predictive power of the DFT+DMFT approach is, obviously, limited by the underlying approximations, namely the neglect of non-local correlations -- blamed for an underestimation of the voltage within DFT+DMFT in the literature \cite{IsaacsMarianetti_Compositional2020} -- and ambiguities concerning the double counting corrections and the quality of the charge density used for generating the non-interacting part of the Hamiltonian.  For example, GGA+DMFT
calculations performed using a Hamiltonian generated with the GGA charge density (or Kohn-Sham Hamiltonian)
without charge self-consistency predict a voltage of 1.49 V, greatly
underestimating the experimental value. In a previous study, it
was noted that non-charge-self-consistent DFT+DMFT calculations
could worsen the predicted voltage compared to conventional DFT \cite{IsaacsMarianetti_Compositional2020}.
Nevertheless, our results suggest that the overall energetics needed for a reasonable description of the voltage is captured in our efficient
charge-self-consistent DFT+DMFT scheme. In future work, we plan to further
investigate the relative effects of non-local correlation vs charge-redistribution using more sophisticated methods such as $GW$+DMFT, which allow for progress both concerning a better description of the quasi-particle band structures and the double counting issue.

\section{Conclusion \label{sec:Conclusion}}

We have performed DFT+DMFT calculations using an efficient charge-self-consistent
scheme to investigate the electronic structure, local properties,
and intercalation voltage of the Li-rich layered sulfides Li$_{x}${[}Li$_{0.33-2y/3}$Ti$_{0.67-y/3}$Fe$_{y}${]}S$_{2}$.
A careful comparison between results using different methods, including spin-averaged and spin-polarized GGA,
and GGA+DMFT was made. Both of the end members, namely the
fully discharged Li$_{1.11}$Ti$_{0.56}$Fe$_{0.33}$S$_{2}$ and the
fully charged compound Li$_{0.11}$Ti$_{0.56}$Fe$_{0.33}$S$_{2}$,
are strongly correlated systems, identified as high-spin Mott-insulator
and low-spin correlated metal, respectively. We have shown that
dynamical correlations originating from Hund's exchange coupling $J_{H}$
are important to describe this class of materials, while the effective local Hubbard
$U$ is largely screened. The impact of $J_{H}$ on the intercalation
voltage may suggest new pathes for designing higher-energy
lithium ion-battery cathode materials. To our knowledge, this is the first demonstration of Hund's physics playing a crucial role in the electrochemcial properties of real-life battery materials. A deeper understanding of these effects will hopefully contribute to paving the way to better battery materials in the future.

\section*{ACKNOWLEDGMENTS}
This work was supported by IDRIS/GENCI Orsay under projet number No.A0110901393.
We  thank the  computer  team at  CPHT for support.
D.D.S. acknowledges funding from Science and Engineering Research Board, Department of Science and Technology, Government of India and Jamsetji Tata Trust.
D.D.S. is also thankful to the Foundation of Ecole Polytechnique for the Gaspard Monge Visiting Professorship.

\appendix

\section{Computational details}

By using DFT within the VASP \cite{KresseFurthmuller_Efficient1996}
code and DFT+DMFT implemented in the DMFTpack software combined with OpenMX
\cite{_DMFTpack,openmx,SimHan_Density2019} and the impurity solver
implemented in Ref. \onlinecite{HauleHaule_Quantum2007}, we have studied
the electronic and magnetic properties of Li$_{x}${[}Li$_{0.33-2y/3}$Ti$_{0.67-y/3}$Fe$_{y}${]}S$_{2}$.
The DFT calculations were performed within the generalized gradient approximation
as parameterized by Perdew, Burke and Ernzerhof (GGA-PBE) \cite{PerdewErnzerhof_Generalized1996}.
The D3 method of Grimme et al. was used for van der Waals corrections \cite{GrimmeKrieg_Consistent2010, GrimmeGoerigk_Effect2011}
 Atomic positions were
relaxed with a force criterion of 1 meV/Å. The lattice constants are
fixed to the experimental values: $a=b=3.54$ Å and $a=b=3.35$ Å
for the discharged ($x=1$) and fully charged ($x=0$) phases, respectively
\cite{SahaTarascon_Exploring2019}. $13\times13\times8$ k-points
were used in the Brillouin zone for the momentum space integrations. 

To describe electronic correlation effects, the GGA$+U$
\cite{AnisimovAndersen_Band1991,AnisimovSawatzky_Densityfunctional1993}
and single-site paramagnetic GGA+DMFT \cite{AnisimovKotliar_Firstprinciples1997, LichtensteinKatsnelson_Initio1998} 
have been employed. The interaction part of the Hamiltonian for the d-shell reads:
\begin{align}
	H_{{\rm int}}= & \frac{1}{2}\sum_{\sigma,m,m^{\prime}}U_{mm^{\prime}}n_{m\sigma}n_{m^{\prime}\bar{\sigma}}\nonumber \\
	& +\frac{1}{2}\sum_{\sigma,m\neq m^{\prime}}(U_{mm^{\prime}}-J_{mm^{\prime}})n_{m\sigma}n_{m^{\prime}\sigma}.\label{eq:Hint}
\end{align}
Here the direct and exchange interaction parameters, $U_{mm^{\prime}}$
and $J_{mm^{\prime}}$, are parameterized by the Slater integrals of the $d$-shell,
namely $F_{0},$ $F_{2}$, and $F_{4}$ \cite{PavariniInstituteforAdvancedSimulation_LDA2011}.
We present our results in terms of the $U=F_{0}$ and $J_{H}=(F_{2}+F_{4})/14$, within the assumption $F_{4}/F_{2}=0.625$.
Unless otherwise stated, calculations were performed with $U=3.0$, $J_{H}=0.9$, and FLLnS double counting : $\Sigma_{DC}^{\rm FLLnS}=U(N-\frac{1}{2})-J(\frac{N}{2}-\frac{1}{2})$ \cite{YlvisakerKoepernik_Anisotropy2009, RyeeHan_Effect2018}.
The charge-density-only GGA$+U$ should be distinguished from other DFT$+U$ flavors such as spin-polarized GGA$+U$ \cite{LiechtensteinZaanen_Densityfunctional1995} and the Dudarev's simplified DFT$+U$ method \cite{DudarevSutton_Electronenergyloss1998, HanYu_LDA2006}. 
The interaction parameters, $U$ and $J_{H}$,
used in this work are consistent with that used in the previous study,
$U_{\text{eff}}=U-J_{H}=2.0$ eV \cite{SahaTarascon_Exploring2019}
and in reasonable range compared to other iron-based compounds
\cite{vanRoekeghemBiermann_Hubbard2016}.
The ferromagnetic ground state, with a lower energy of 3 meV than the antiferromagnetic state at $U=3.0$ and $J_{H}=0.9$ eV, is assumed for various $U$ and $J_{H}$ parameters.

In our single site DMFT calculation, a natural atomic orbitals projector
onto to Fe-$d$ orbitals with an energy window of $W=[-6,5]$ eV containing
Fe-$d$ and S-$p$ orbitals has been employed \cite{SimHan_Density2019}.
The self-energy is decomposed into three $10\times10$ matrices corresponding
to three inequivalent Fe atoms in the unit cell, i.e., $\Sigma(i\omega_{n})=\oplus_{i=1}^{3}\Sigma_{{\rm Fe(i)}}(i\omega_{n})$.
$\Sigma_{{\rm Fe(i)}}(i\omega_{n})$ is determined from the fictitious
impurity problem with self-consistency condition. The impurity problems
are solved by employing a hybridization expansion continuous-time
quantum Monte Carlo (CT-QMC) \cite{WernerMillis_ContinuousTime2006,WernerMillis_Hybridization2006}
algorithm implemented in Ref. \cite{HauleHaule_Quantum2007}. The
self-energy in the real frequency domain is obtained from the Matsubara
self-energy by analytic continuation using the maximum quantum entropy
method \cite{SimHan_Maximum2018}, extending the maximum entropy method
to matrix valued Green's functions \cite{JarrellGubernatis_Bayesian1996,GunnarssonSangiovanni_Analytical2010a}. 

\section{On-site Coulomb energy in DFT+$U$ and DFT+DMFT\label{sec:On-site-Coulomb-energy}}

In this section we calculate the total energy contribution from the
on-site interaction $\Delta E_{U}=\braket{H_{{\rm int}}}-E_{{\rm DC}}$.
For simplicity, in this section the interaction $H_{{\rm int}}$
is assumed to be of Slater-Kanamori form with $U_{SK}=U+8J_{H}/7,$ $J_{SK}=5J_{H}/7$,
and $U'_{SK}=U_{SK}-2J_{SK}$ \cite{PavariniInstituteforAdvancedSimulation_LDA2011,IsaacsMarianetti_Compositional2020}. 

\paragraph{DFT$+U$}

$E_{{\rm int}}^{U}=\frac{U}{2}\sum_{(\alpha\sigma)\neq(\beta\sigma\prime)}n_{\alpha\sigma}n_{\beta\sigma^{\prime}}-\frac{J_{H}}{2}\sum_{\sigma}\sum_{\alpha\neq\beta}n_{\alpha\sigma}n_{\beta\sigma}$
and the double counting term is $E^{{\rm DC}}_{{\rm FLLnS}}=\frac{U}{2}N(N-1)-\frac{J}{2}\sum_{\sigma}\frac{N}{2}(\frac{N}{2}-1)$,
where $n_{\alpha\sigma}$ is an eigenvalue of the occupation number
matrix with orbital index $\alpha$ and spin $\sigma$. $N=\sum_{\sigma}N^{\sigma}=\sum_{\sigma}\sum_{\alpha}n_{\alpha\sigma}$.
The Coulomb energy correction $\Delta E^{U}=E_{int}^{U}-E_{{\rm DC}}$
can be written as follows:
\[
\Delta E^{U}=\frac{1}{2}(U-J)\sum_{\alpha\sigma}(n_{\alpha\sigma}-n_{\alpha\sigma}^{2})-\frac{J}{4}M^{2},
\]
 where $M=N_{\uparrow}-N_{\downarrow}$ is the magnetic moment of
the localized $d$-orbitals. We note that the first term is obtained
from the double counting energy $E_{{\rm DC}}^{{\rm FLL}}$ \cite{HanYu_LDA2006}.
The second
term imposes an energy gain due to the finite magnetic moment $M$, leading to a magnetic
polarization, which is not included in the non-spin polarized GGA
exchange-correlation functional.

In DMFT, the Coulomb interaction energy is beyond the HF approximation.
$E_{{\rm int}}^{{\rm DMFT}}=\frac{U}{2}\braket{\hat{N}^{2}-\hat{N}}-\frac{J_{H}}{2}\sum_{\sigma}\braket{\hat{N_{\sigma}}^{2}-\hat{N_{\sigma}}}$.
After some algebra, using $E_{\text{DC}}^{\text{FLLnS}}$, we obtain
\[
\Delta E^{{\rm DMFT}}=\frac{1}{2}(U-J)\sum_{\alpha\sigma}(\braket{\delta N^{2}})-\frac{J}{4}M_{{\rm eff}}^{2},
\]
where $\braket{\delta N^{2}}=\braket{N^{2}}-\braket{N}^{2}$ and $M_{{\rm eff}}=(N^{2}-4\braket{N_{\uparrow}N_{\downarrow}})^{1/2}$.
We can see that within the HF approximation, $\braket{N_{\uparrow}N_{\downarrow}}\approx N_{\uparrow}N_{\downarrow}$,
$M_{{\rm eff}}$ is equivalent to the conventional magnetization,
namely $M_{{\rm eff}}=N_{\uparrow}-N_{\downarrow}$.

\paragraph{Atomic limit}

Three different configurations, namely, $d^{6}$(HS), $d^{6}$(LS), and $d^{5}$(LS) are concerned. Here, the $d^{6}$(HS)
and $d^{5}$(LS) configurations are expected as ground state configurations of Fe in LTFS$_{1}$ and LTFS$_{0}$, respectively.
Assuming FLL-nS double counting, one can see that $\Delta E_{U}$ can be expressed
as:
\begin{align}
\Delta E_{U}\approx\begin{cases}
-4J_{H}+2\bar{\Delta} & d^{6}(HS)\\
1.71J_{H} & d^{6}(LS)\\
0.89J_{H} & d^{5}(LS)
\end{cases}.
\end{align}
Given these expressions, we can estimate $J_{H}$ dependence of the
Coulomb energy contribution to the voltage for different atomic configurations.
Interestingly, our DFT+DMFT results show  similar trends for
the transition from the  $d^{6}$ to the $d^{5}$(LS) configuration:
\begin{align}
V_{LS}=E(d^{6})-E(d^{5})\propto & \begin{cases}
4.89J_{H}, & \text{for }J_{H}/\Delta>0.35\\
-0.82J_{H}, & \text{otherwise}.
\end{cases}
\end{align}

\bibliography{LTFS}

\end{document}